\documentclass{aa}  
\usepackage{graphicx}
\usepackage{txfonts}
\begin{document}

   \title{The ASTRODEEP Frontier Fields Catalogues: II - Photometric redshifts and rest-frame properties in Abell-2744 and MACS-J0416}

\author{M. ~Castellano \inst{1}
\and R. ~Amor\'{\i}n \inst{1}
\and E. ~Merlin \inst{1}
\and A. ~Fontana \inst{1}
\and R. ~J. ~McLure \inst{2}
\and E. ~M\'armol-Queralt\'o \inst{2}
\and A. ~Mortlock \inst{2}
\and S. ~Parsa \inst{2}
\and J. ~S. ~Dunlop \inst{2}
\and D. ~Elbaz \inst{3}
\and I. ~Balestra \inst{4}
\and A. ~Boucaud \inst{5,18}
\and N. ~Bourne \inst{2}
\and K. ~Boutsia \inst{1}
\and G. ~Brammer \inst{6}
\and V. ~A. ~Bruce \inst{2}
\and F. ~Buitrago \inst{2,14,15}
\and P. ~Capak \inst{7}
\and N. ~Cappelluti \inst{8}
\and L. ~Ciesla \inst{3}
\and A. ~Comastri \inst{8}
\and F. ~Cullen \inst{2}
\and S. ~Derriere \inst{9}
\and S. ~M. ~Faber \inst{10}
\and E. ~Giallongo  \inst{1}
\and A. ~Grazian  \inst{1}
\and C. ~Grillo \inst{11}
\and A. ~Mercurio \inst{12}
\and M.~J.~Micha{\l}owski \inst{2}
\and M. ~Nonino \inst{4}
\and D. ~Paris  \inst{1}
\and L. ~Pentericci \inst{1}
\and S. ~Pilo  \inst{1}
\and P. ~Rosati \inst{13}
\and P. ~Santini \inst{1} 
\and C. ~Schreiber \inst{3}
\and X. ~Shu  \inst{3,16} 
\and T. ~Wang  \inst{3,17}
}
\institute{INAF - Osservatorio Astronomico di Roma, Via Frascati 33, I - 00040 Monte Porzio Catone (RM), Italy \email{marco.castellano\char64oa-roma.inaf.it}\label{inst1}
\and SUPA\thanks{Scottish Universities Physics Alliance}, Institute for Astronomy, University of Edinburgh, Royal Observatory, Edinburgh, EH9 3HJ, U.K. \label{inst2}
\and Laboratoire AIM-Paris-Saclay, CEA/DSM/Irfu - CNRS - Universit\'e Paris Diderot, CEA-Saclay, pt courrier 131, F-91191 Gif-sur-Yvette, France \label{inst3}
\and INAF - Osservatorio Astronomico di Trieste, Via G. B. Tiepolo 11, I-34143 Trieste, Italy \label{inst4} 
\and Institut d'Astrophysique Spatiale, CNRS, UMR 8617, Univ. Paris-Sud, Universit\'{e}
Paris-Saclay, IAS, b\u{a}t. 121, Univ. Paris-Sud, 91405 Orsay, France \label{inst5}
\and Space Telescope Science Institute, 3700 San Martin Drive, Baltimore, MD 21218, USA \label{inst6}
\and Spitzer Science Center, 314-6 Caltech, Pasadena, CA 91125, USA \label{inst7} 
\and INAF - Osservatorio Astronomico di Bologna, Via Ranzani 1, I - 40127, Bologna, Italy \label{inst8}
\and Observatoire astronomique de Strasbourg, Universit\'e de Strasbourg,
CNRS, UMR 7550, 11 rue de l'Universit\'e, F-67000 Strasbourg, France\label{inst9} 
\and UCO/Lick Observatory, University of California, 1156 High Street, Santa Cruz, CA 95064, USA \label{inst10} 
\and Dark Cosmology Centre, Niels Bohr Institute, University of Copenhagen, Juliane Maries Vej 30, 2100 Copenhagen, Denmark \label{inst11} 
\and INAF - Osservatorio Astronomico di Capodimonte, via Moiariello 16 80131 Napoli, Italy \label{inst12} 
\and Dipartimento di Fisica e Scienze della Terra, Università degli Studi di Ferrara, via Saragat 1, 44122 Ferrara, Italy \label{inst13} 
\and Instituto de Astrof\'{\i}sica e Ci\^{e}ncias do Espa\c{c}o, Universidade de Lisboa, OAL, Tapada da Ajuda, PT1349-018 Lisbon, Portugal \label{inst14} 
\and Departamento de F\`{i}sica, Faculdade de Ci\^{e}ncias, Universidade de Lisboa, Edif\`{i}cio C8, Campo Grande, PT1749-016 Lisbon, Portugal \label{inst15} 
\and Department of Physics, Anhui Normal University, Wuhu, Anhui, 241000, China \label{inst16} 
\and School of Astronomy and Astrophysics, Nanjing University, Nanjing, 210093, China \label{inst17} 
\and Sorbonne Universit\'es, UPMC Univ Paris 6 et CNRS, UMR 7095,  Institut d'Astrophysique de Paris, 98 bis bd Arago, 75014 Paris, France. \label{inst18}
}

\authorrunning{M. Castellano, R. Amorin, E. Merlin et al.}
\titlerunning{\textsc{ASTRODEEP Frontier Fields Catalogues II}}

   \date{}

 %
  \abstract
   {}
   {We present the first public release of photometric redshifts, galaxy rest-frame properties and associated magnification values in the cluster and parallel pointings of the first two Frontier Fields, Abell-2744 and MACS-J0416. The released catalogues aim at providing a reference for future investigations of the extragalactic populations in these legacy fields: from lensed high-redshift galaxies to cluster members themselves.}
   {We exploit a multi-wavelength catalogue ranging from HST to ground-based K and Spitzer IRAC which is specifically designed to enable detection and measurement of accurate fluxes in crowded cluster regions. The multi-band information is used to derive photometric redshifts and physical properties of sources detected either in the H-band image alone or from a stack of four WFC3 bands. To minimize systematics median photometric redshifts are assembled from six different approaches to photo-z estimates. Their reliability is assessed through a comparison with available spectroscopic samples. State of the art lensing models are used to derive magnification values on an object-by-object basis by taking into account sources positions and redshifts.}
   {We show that photometric redshifts reach a remarkable $\sim$3-5\% accuracy. After accounting for magnification the H band number counts are found in agreement at bright magnitudes with number counts from the CANDELS fields, while extending the presently available samples to galaxies intrinsically as faint as H160$\sim$32-33 thanks to strong gravitational lensing. The Frontier Fields allow to probe the galaxy stellar mass distribution at 0.5-1.5 dex lower masses, depending on magnification, with respect to extragalactic wide fields, including sources at $M_{star} \sim 10^7$-$10^8 M_{\odot}$ at z$>$5. Similarly, they allow the detection of objects with intrinsic SFRs $>$1dex lower than in the CANDELS fields reaching 0.1-1 $M_{\odot}/yr$ at z$\sim$6-10.}
   {}


   \maketitle

\section{Introduction}

The use of photometric redshifts and SED-fitting techniques is acquiring an ever increasing importance for investigating the properties of extragalactic populations where spectroscopic studies of large flux-limited samples are beyond reach of current instrumentation. To this aim, large efforts have been spent to assemble determinations of both photo-zs and galaxy rest-frame properties from the available multi-band datasets of deep field surveys, such as GOODS \citep{Grazian2006}, COSMOS \citep{Ilbert2009}, CANDELS \citep{Dahlen2013} and 3D-HST \citep{Skelton2014}. The relevance of these analysis is well shown by the emerging collaborative efforts combining different codes and techniques to smooth out possible systematics in the computation of robust photo-zs and rest-frame properties \citep[e.g.][]{Santini2015,Mobasher2015}.

Accurate estimates of photometric redshifts and galaxy properties are today the missing ingredient for exploiting the Frontier Fields (FF) survey, an HST observing program targeting six galaxy clusters fields, and six parallel ``blank'' fields at depths comparable to the Hubble Ultra Deep Field one. Thanks to the  magnification by the foreground galaxy clusters the FF survey enables the detection of galaxies as intrinsically faint as future JWST targets while also reducing cosmic variance effects in the study of ultra-faint galaxy populations thanks to the independent pointings. The FF survey hold promise for becoming a milestone in extragalactic studies in the forthcoming years.

In this paper we present a public release of photometric redshifts and rest-frame galaxy properties from multi-wavelength photometry of the Frontier Fields Abell-2744 (A2744 hereafter) and MACS-J0416 (M0416 hereafter) cluster and parallel fields including both HST and deep K band and Spitzer information. A detailed description of the dataset and photometric measurements is presented in a companion paper by Merlin et al. (M16 hereafter). The multi-band and photometric redshift catalogues of the FF have been developed in the context of the European  FP7-Space project ASTRODEEP\footnote{\textsc{Astrodeep} is a coordinated and comprehensive program of i)
algorithm/software development and testing; ii) data reduction/release,
and iii) scientific data validation/analysis of the deepest
multi-wavelength cosmic surveys. For more information, visit
\textit{http://astrodeep.eu}}. The plan of the paper is the following: in Sect.~\ref{sect_MW} we briefly describe the available photometric and spectroscopic data and the catalogue assembly procedure from M16. Sect.~\ref{sect_PHOTOZ} will introduce our procedure for estimating photometric redshifts and an evaluation of their accuracy. The determination of magnification values on an object-by-object basis and the resulting de-magnified number counts are discussed in Sect.~\ref{sect_NCOUNTS}, while de-magnified stellar masses and star-formation rates are presented in Sect.~\ref{sect_RESTPROP}. Finally a summary of the work is given in Sect.~\ref{sect_SUMMARY} and a description of the publicly available dataset\footnote{Download: http://www.astrodeep.eu/frontier-fields-download/;\\ Catalogue interface: http://astrodeep.u-strasbg.fr/ff/index.html} is included in Sect.~\ref{sect_APP_CATAL}.

Throughout the paper, observed and rest--frame magnitudes are in
the AB system, and we adopt the $\Lambda$-CDM concordance model ($H_0=70km/s/Mpc$, $\Omega_M=0.3$, and $\Omega_{\Lambda}=0.7$).

\section{Multi-wavelength catalogues}\label{sect_MW}

A detailed description of the dataset and of the catalogue assembly strategy is provided in M16; we summarise here the information most relevant for the work we present in this paper.

\subsection{Dataset}
\begin{figure}[!ht]
   \centering
   \includegraphics[width=7cm, angle=-90]{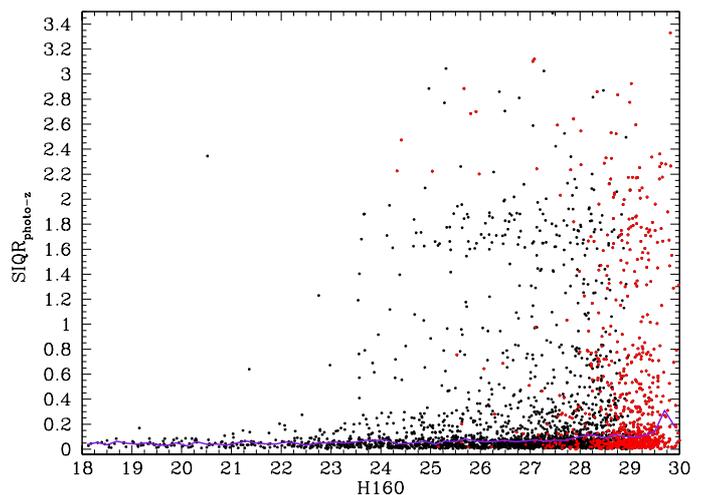}
   \caption{Semi-interquartile range of the six different photo-z estimates as a function of the H-band magnitude (or upper limit) for H-detected (black circles) and IR-detected (red) sources in the Abell-2744 cluster field. The median SIQR as a function of magnitude is shown as a purple line.}\label{fig_siqr}%
    \end{figure}

The A2744 and M0416 are the first 2 of a total of 6 twin fields, observed by HST in parallel (i.e. the cluster pointing together with a "blank" parallel pointing) in 3 optical and 4 near-infrared bands: F435W, F606W and F814W (ACS);  F105W, F125W, F140W and F160W (WFC3). The HST bands have a typical 5$\sigma$ depth in the range 28.5-29.0 AB in 2 PSF-FWHM apertures. Along with the 7 HST bands we include in each field the publicly available Hawk-I@VLT \textit{Ks} images from ESO Programme 092.A-0472\footnote{P.I. G. Brammer, http://gbrammer.github.io/HAWKI-FF/} ($\sim$26.2 at 5$\sigma$), and the IRAC 3.6 and 4.5 $\mu$m data acquired under DD time and, in the case of M0416, Cycle-8 program iCLASH (80168) ($\sim$25 AB at 5$\sigma$).

To fully exploit the depth of the images and to detect outshined faint sources, we developed a procedure to remove the foreground light of bright cluster sources and the intra-cluster light (ICL). We start with the H160 image applying the following procedure:

\begin{figure*}[ht]
   \centering
   \includegraphics[width=6.5cm, angle=-90]{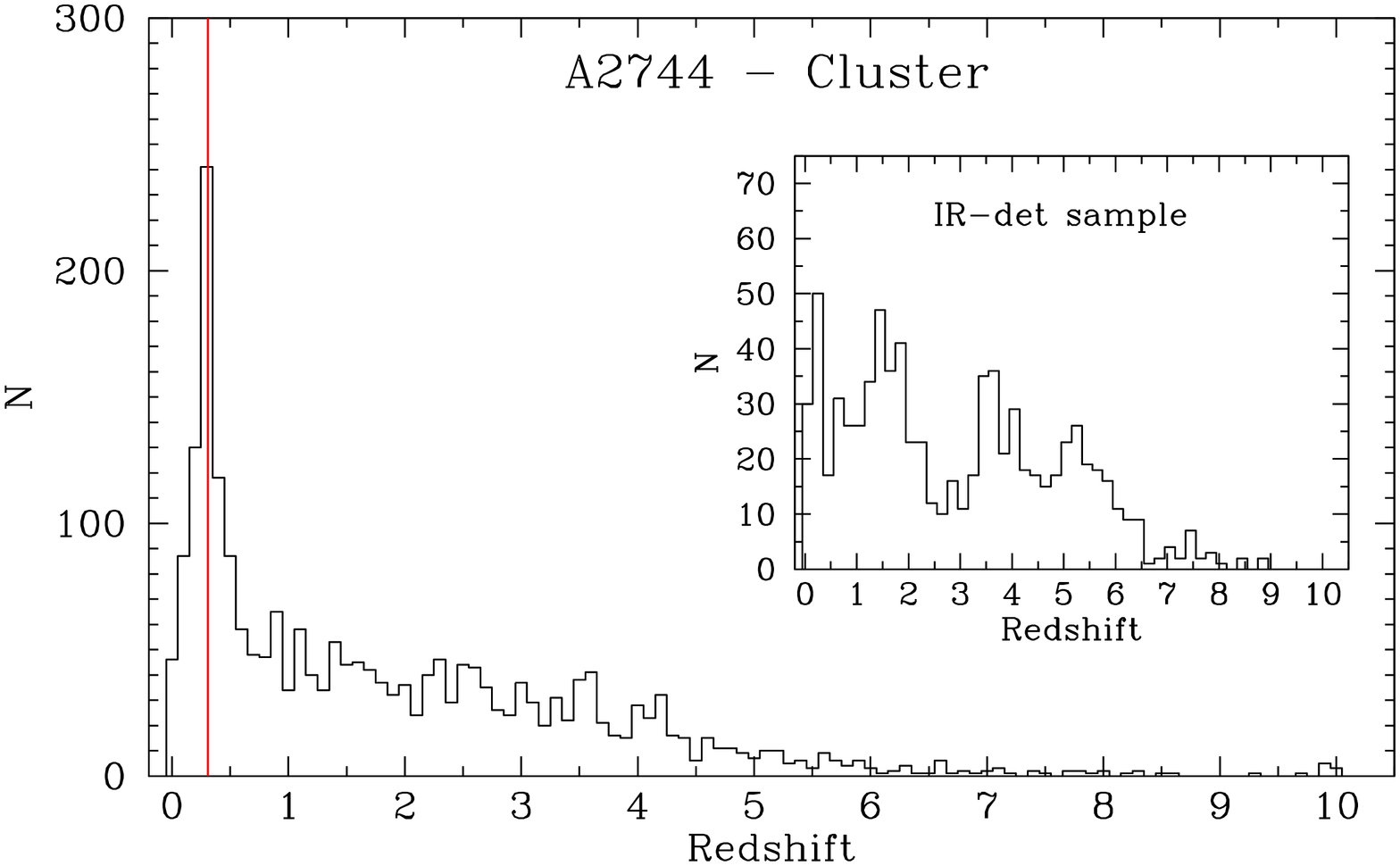}
   \includegraphics[width=6.5cm, angle=-90]{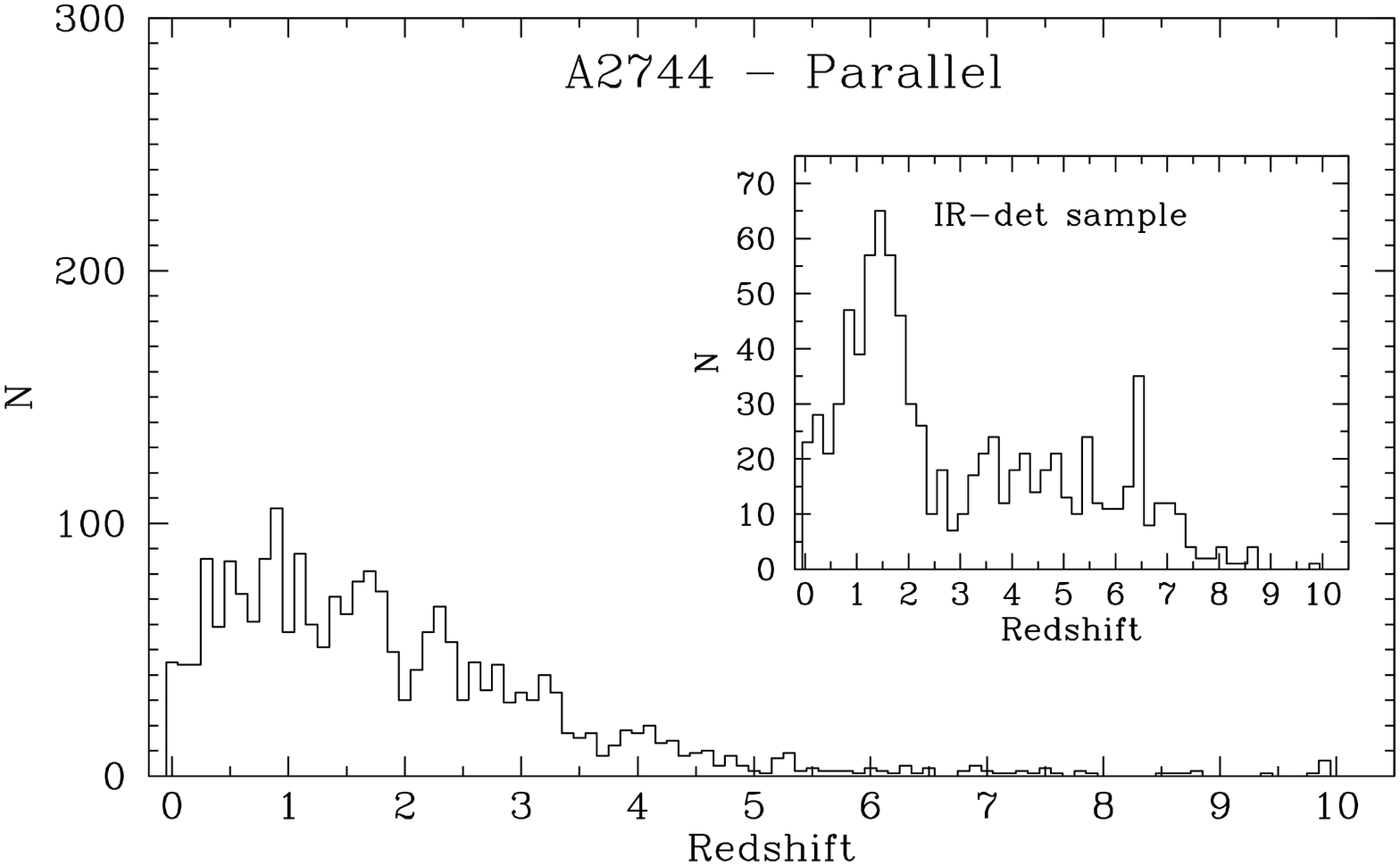}
   \includegraphics[width=6.5cm, angle=-90]{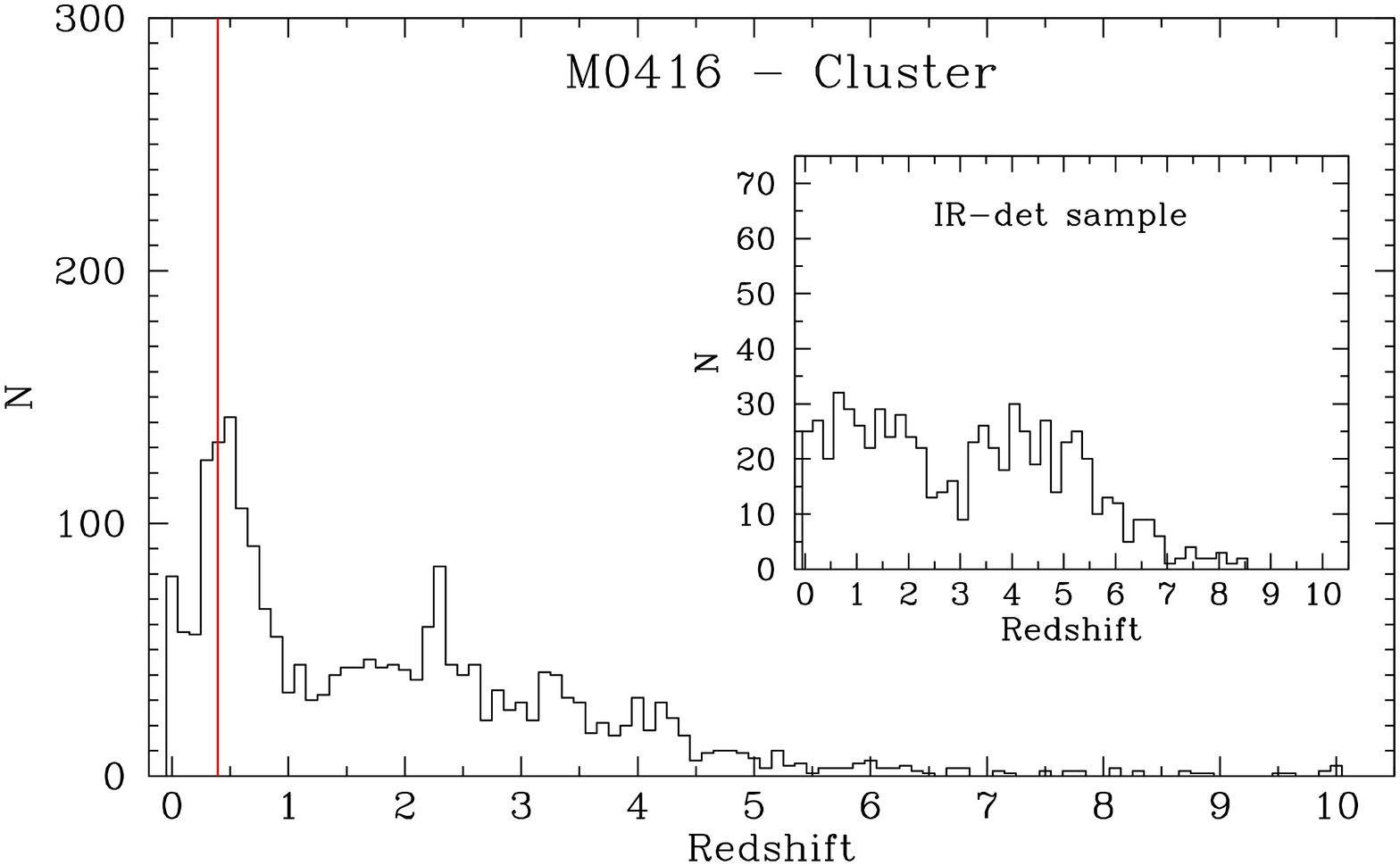}
   \includegraphics[width=6.5cm, angle=-90]{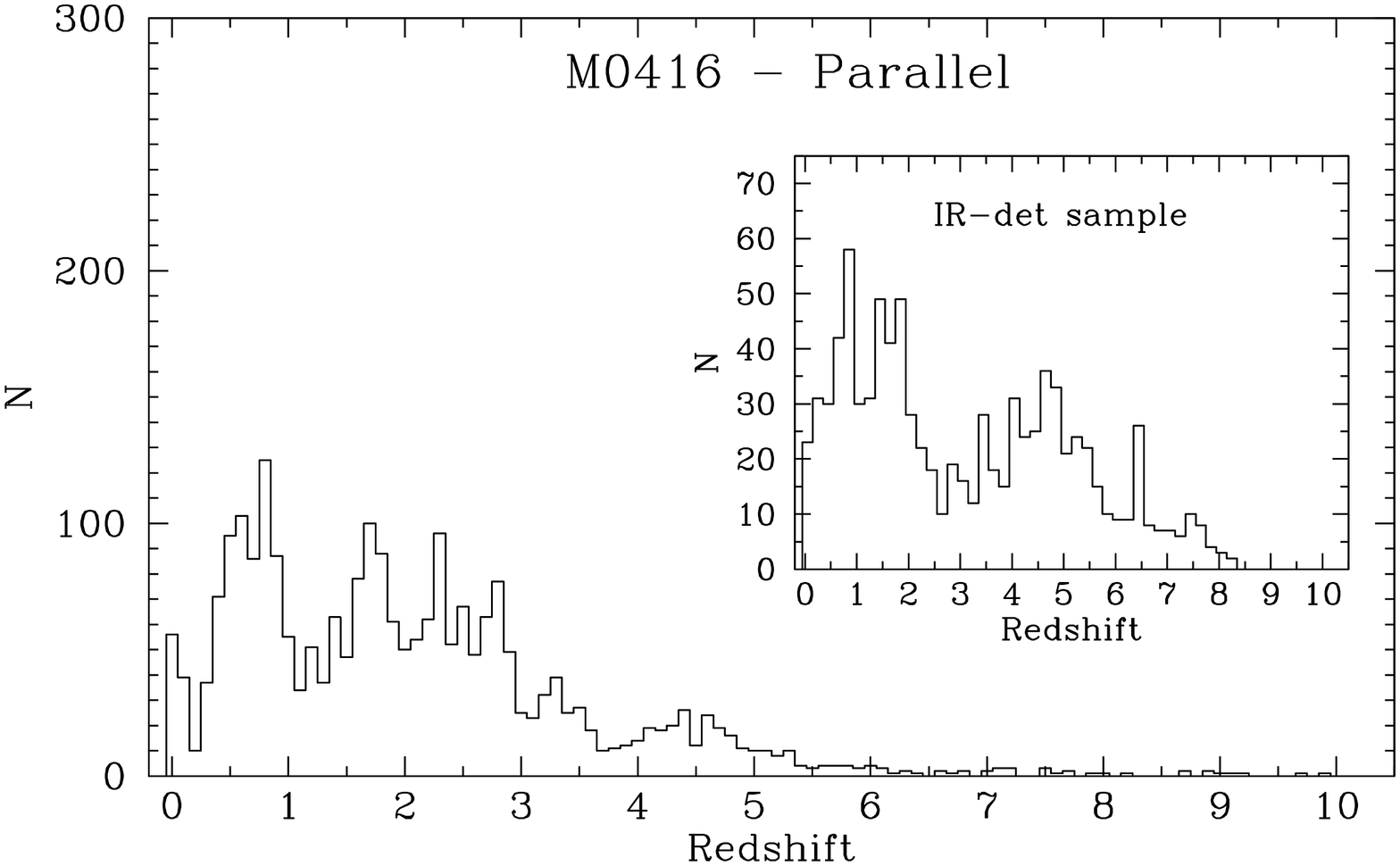}
   \caption{Photometric redshift distribution of H-detected catalogues in, from top to bottom, left to right: Abell-2744 Cluster, Abell-2744 Parallel, MACS-0416 Cluster, MACS-0416 Parallel. Inset plots show the distribution of the additional IR-detected samples. Vertical red lines mark the redshift of the lensing clusters.}
              \label{fig_zdistr}%
    \end{figure*}
1) a first raw estimate of the ICL component is obtained by masking S/N$>$8 pixels in the original H160 image and fitting the ICL light with \verb|Galfit|  \citep{Peng2010}, using one or more Ferrer profiles \citep[see][]{Giallongo2014}. The best-fit model is then subtracted from the original image. 2) on the ICL-subtracted H160 image, we use \verb|Galapagos| \citep{Barden2012} to obtain a single Sersic fit of the brightest cluster members, typically mag$<$19 galaxies close to the cluster center. 3) we then progressively refine the fit for such objects by adding a second, ``bulgy'' component and fitting again with \verb|Galfit|, leaving the structural parameters of the galaxies free to adjust; if necessary, we iterate the procedure adding further components until a satisfying residual is obtained;
4) having obtained the best fit for the galaxies, we keep them fixed and fit again the ICL with \verb|Galfit| on the original image, letting its parameters free to adjust; then we obtain a ``final residual'' image by subtracting this final ICL model and the bright galaxy models from the image; 5) finally, we create a median filtered version of the residual images over a 1$\times$1 arcsec box. To avoid affecting signal from faint objects we exclude from the computation all pixels at $>$1$\sigma$ above zero counts and their nearest neighbors. We obtain the final processed frame by subtracting the resulting median-filtered image from the ``final residual'' one, thus smoothing out local fluctuations and \verb|Galfit| residuals and allowing for a more efficient detection.

We subtract ICL and bright sources from the other HST bands using the final fitting parameters of the nearest redder band as starting guesses and simultaneously fitting all the components at once.  Our  catalogue is extracted by performing the detection on the final processed H160 image with \verb|SExtractor| \citep{Bertin1996} using a customized version of the HOT+COLD approach \citep{Galametz2013,Guo2013}. The resulting 90\% detection completeness limits for point sources and disk-like galaxies as estimated through simulations are at H160$\sim$27.75 and H160$\sim$27.25 respectively (M16). We extract fluxes from the other HST bands with \verb|SExtractor| in dual mode after having  PSF-matched them to the H160 PSF through appropriate convolution kernels derived from bright unsaturated stars. Total fluxes in the detection band are estimated from SExtractor \verb|FLUX_AUTO|. Total fluxes in the other HST bands are computed by scaling the total flux in the detection band on the basis of the relevant isophotal colours computed from SExtractor \verb|FLUX_ISO| values. K and IRAC photometry is obtained via a template-fitting technique with \verb|T-PHOT| \citep{Merlin2015} using galaxy shapes in the detection band as ``prior'' information. \verb|T-PHOT| allows us to fit ``real'' sources together with analytical models, therefore we used the detected H160 catalog plus the bright source models as priors. Before the fit, measurement images are processed re-estimating the background and the RMS via injection of fake PSF-shaped sources in void regions. Also, a local background subtraction is performed during the fit, allowing for a better estimation of the flux for objects falling within the halos of bright sources. All fluxes are corrected for galactic extinction derived from \citet{Schlegel1998} dust emission maps.
\begin{figure*}[ht]
   \centering
   \includegraphics[width=9cm, ]{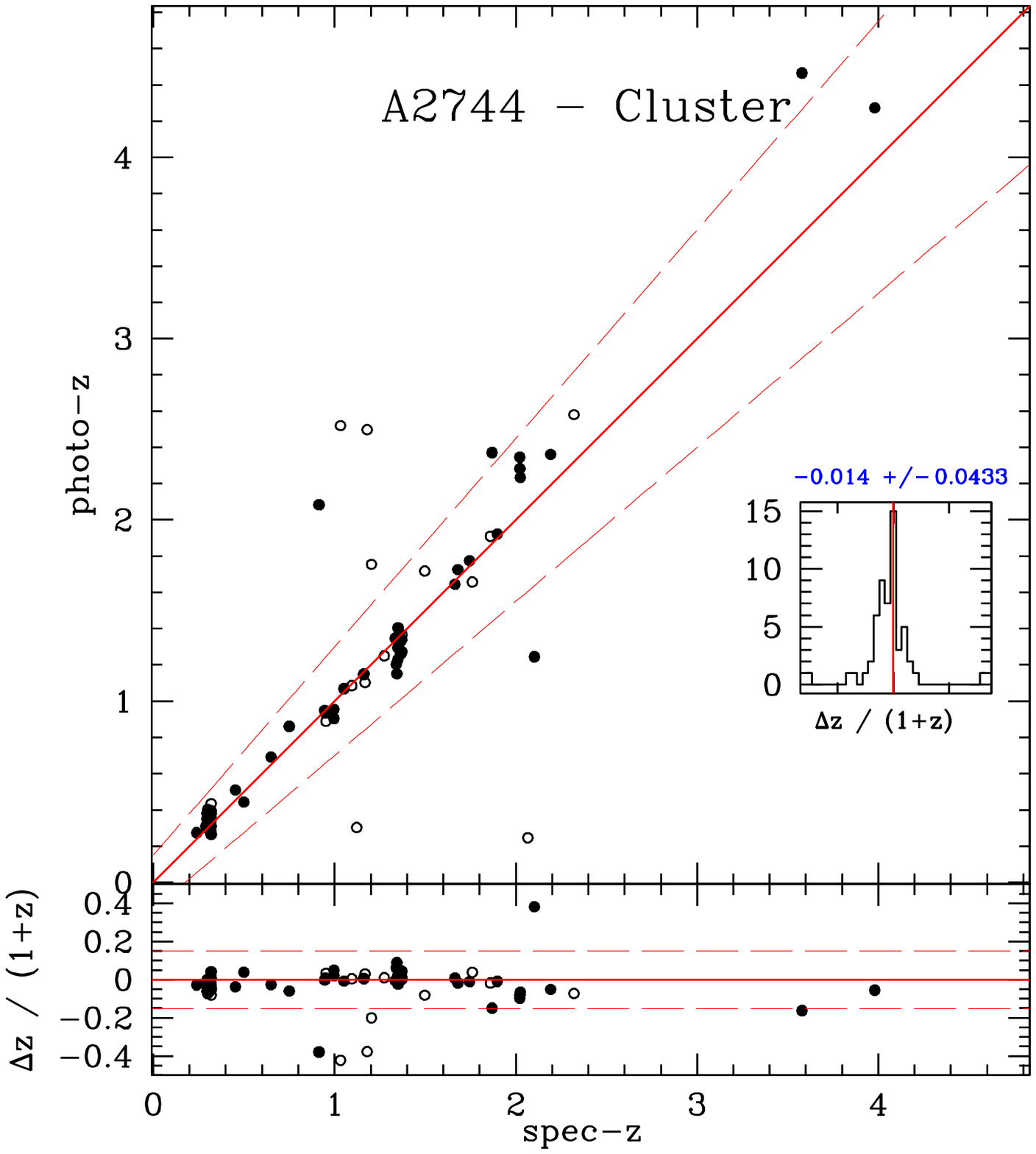}
   \includegraphics[width=9cm, ]{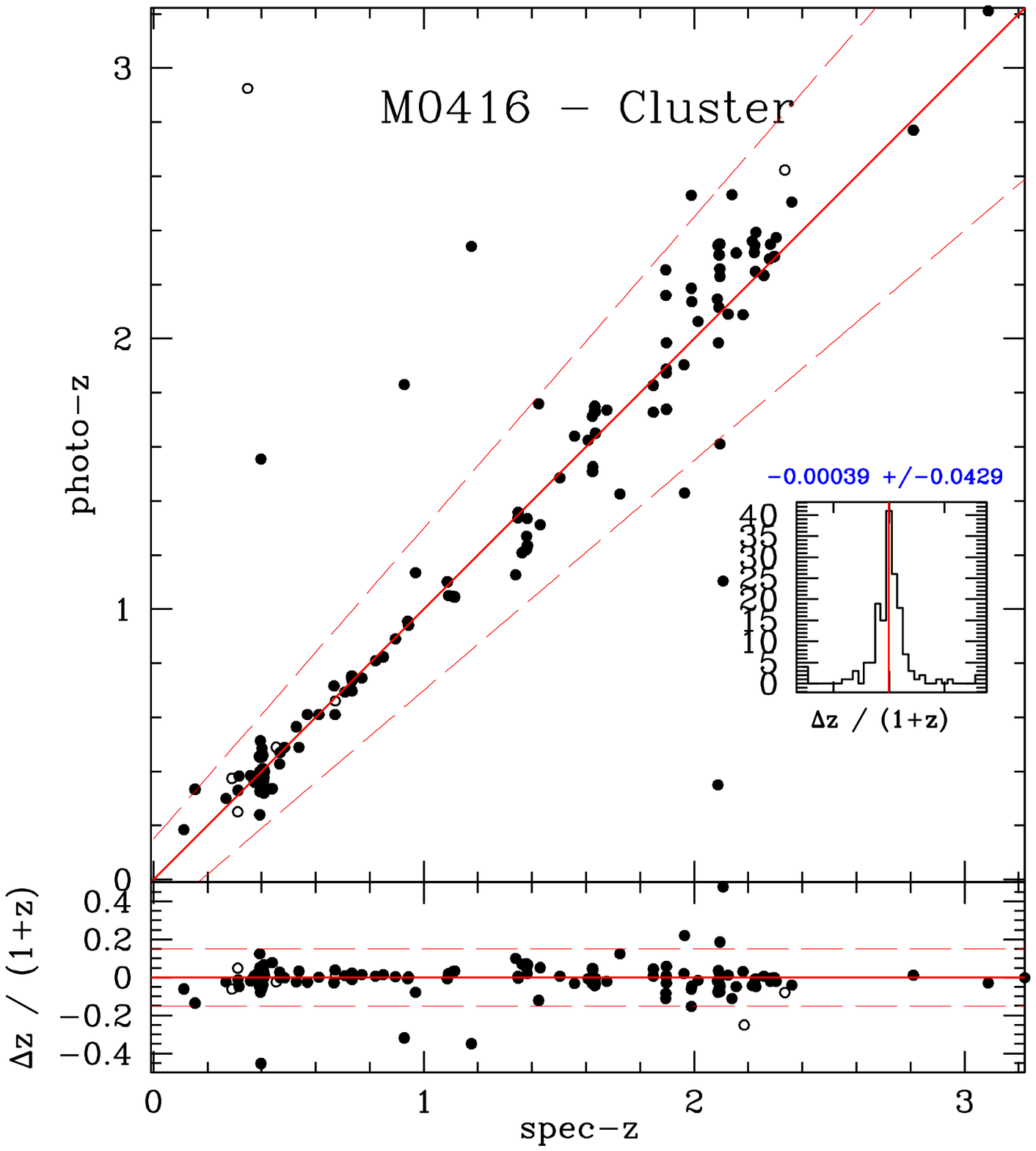}
   \caption{Comparison between photometric and spectroscopic redshifts in the Abell-2744 (left) and MACS-0416 (right) clusters. Filled circles represent best quality spectroscopic redshifts used to compute the photometric redshift accuracy reported in Table~\ref{tab1}, empty circles objects with ``reliable'' redshift from the GLASS sample (quality flag = 3). Lower panels show the $\Delta z/(1+z_{spec})=(z_{spec}-z_{phot})/(1+z_{spec})$ as a function of $z_{spec}$. The inset in each of the upper panels present the relevant distribution of $\Delta z/(1+z_{spec})$ with its average and rms after excluding $|\Delta z /(1+z_{spec})| > 0.15$ outliers as discussed in the text.  Red dashed lines in both panels enclose the $|\Delta z/(1+z_{spec})| \leq 0.15$ region.}
              \label{fig_zspezbest}%
\end{figure*}
\begin{table*}[!ht]
\caption{Photometric redshifts accuracy}             
\label{tabPHZ}     
\centering                         
\begin{tabular}{c c c c c}        
\hline\hline                 
Field & Spec. sample & N. outliers (fraction) & $\langle\Delta z/(1+z)\rangle$ & $\sigma_{\Delta z/(1+z)}$  \\    
\hline                       
   A2744-Cl & 54 & 4 (7.3\%) &  -0.0140  &  0.043\\     
   A2744-Par &9 & 0 (0\%) & 0.0004 & 0.056  \\
   M0416-Cl &155 & 10 (6.5\%) &-0.0004 & 0.043 \\
   M0416-Par &33 & 3 (9\%) & -0.0299 & 0.0362  \\
\hline                                 
\end{tabular}
\end{table*}\label{tab1}
To include all faint sources of potential interest we also perform an additional detection using \verb|SExtractor| with the same parameter set on a weighted average of the processed Y105, J125, JH140 and H160 images, and derive photometry in the other bands in the same way as for the H-detected sample. The final list of detected sources comprises the main H-detected catalogue plus all those IR-detected ones whose segmentation does not overlap with any pixel belonging to H-detected objects according to the relevant segmentation maps. The final catalogues contain 10 band information for 2596 (H-detected)+976 (IR-detected) sources in A2744-Cluster, 2325+1086 in A2744-Parallel, 2556+832 in M0416-Cluster and 2581+1152 in M0416-Parallel.

\subsection{Spectroscopic samples}

We look for counterparts of our sources in available spectroscopic samples by performing a cross-correlation within 1 arcsec radius. We consider the following public datasets: \citet{Owers2011} (objects with quality flag Q=4 or higher) and the arcs from Richard et al. (in prep.) for Abell-2744; \citet{Ebeling2014}, and the arcs from \citet{Grillo2015} and \citet{Christensen2012} for MACS-0416. For both A2744 and M0416 cluster we include redshifts with quality flag Q=3 and Q=4 from the Grism Lens-Amplified Survey from Space \citep[GLASS; GO-13459; PI: Treu, Hoag et al. 2015, in preparation;][]{Treu2015, Wang2015}.
Objects having a positive match with reliable public samples are assigned the measured spectroscopic redshift in our catalogues. To assess photo-z reliability we also match our catalogues with the M0416 proprietary redshift from the CLASH-VLT survey \citep[ESO Large Programme 186.A-0.798, PI: Rosati, ][]{Rosati2014,Balestra2015}.
The final samples include 86, 10, 194 public spectroscopic redshifts in Abell-2744 cluster field, Abell-2744 parallel field and MACS0416 cluster field respectively. No public spectroscopic redshifts are found in the MACS0416 parallel field. Thanks to the addition of the aforementioned proprietary data we reach a total of 207 and 33 spectroscopically confirmed objects in the MACS0416 cluster and MACS0416 parallel fields respectively.

\section{Photometric redshifts}\label{sect_PHOTOZ}

We measure photometric redshifts for all the sources in our catalogues with six different techniques: 1) \verb|OAR|; 2) \verb|McLure|; 3) \verb|Mortlock|; 4) \verb|Parsa|; 5) \verb|Marmol-Queralto-1|; 6) \verb|Marmol-Queralto-2|. The \verb|OAR| photometric redshifts are obtained with the \verb|zphot.exe| code following the well-tested procedure described in  \citet{Fontana2000} and \citet{Grazian2006} \citep[see also][]{Dahlen2013,Santini2015}.  Best-fit photo-zs are obtained through a $\chi^2$ minimization over the observed HST+IR bands using SED templates from \verb|PEGASE| 2.0 \citep{Fioc1997} at 0.0$<$z$<$10.0. We set flux=0 in place of negative values, and a minimum allowed photometric uncertainty corresponding to 0.05 mags for the HST and $Ks$ bands and to 0.1 mags for the IRAC bands: errors smaller than these values are replaced by the minimum allowed uncertainty. The \verb|Parsa| and \verb|Mortlock| runs both use the publicly available \verb|Le Phare| code \citep{Arnouts1999b}, and employ the \verb|PEGASE| and zCOSMOS \citep{Ilbert2006} template sets respectively. The \verb|Marmol-Queralto| runs both utilize the publicly available \verb|EAZY| code \citep{Brammer2008} and employ the PCA \citep[built following][]{Blanton2007} and \verb|PEGASE| template sets respectively. The \verb|McLure| run is based on his own proprietary code, as described in \citet{McLure2011}, which employs \citet{Bruzual2003} templates. All of the photometric redshift runs with the exception of the \verb|OAR| one applied adjustments to the photometric zeropoints and the \verb|McLure|, \verb|Parsa| and \verb|Mortlock| runs included strong nebular emission lines in the SED fits. To minimize systematics due to the use of a single approach we set as reference photo-z for each object the median value from the six available estimates. In Fig.~\ref{fig_siqr} we show the uncertainty (SIQR, semi-interquartile range) on the median photo-z as a function of the observed H-band magnitude for sources in the Abell-2744 cluster field. The typical SIQR (purple line in Fig.~\ref{fig_siqr}) ranges from 0.05 at bright magnitudes to 0.3 for sources at $H>29$. The fraction of sources with highly uncertain median photo-z (SIQR>1) is below $\sim$10\% up to H$\sim$26.0 and reaches $\sim$20\% at the faintest magnitudes. Similar results are found in the other fields under analysis.

\begin{figure}[!ht]
   \centering
    \includegraphics[width=9cm]{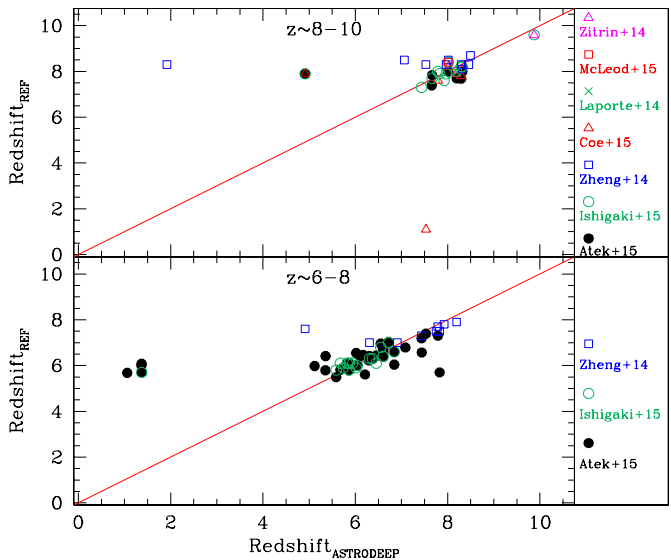}
   \caption{Comparison between photometric redshifts from our catalogues and those from previous papers on high-redshift LBG samples.}\label{fig_highz}%
    \end{figure}
\begin{figure}[!ht]
   \centering
   \includegraphics[width=6.5cm, angle=-90]{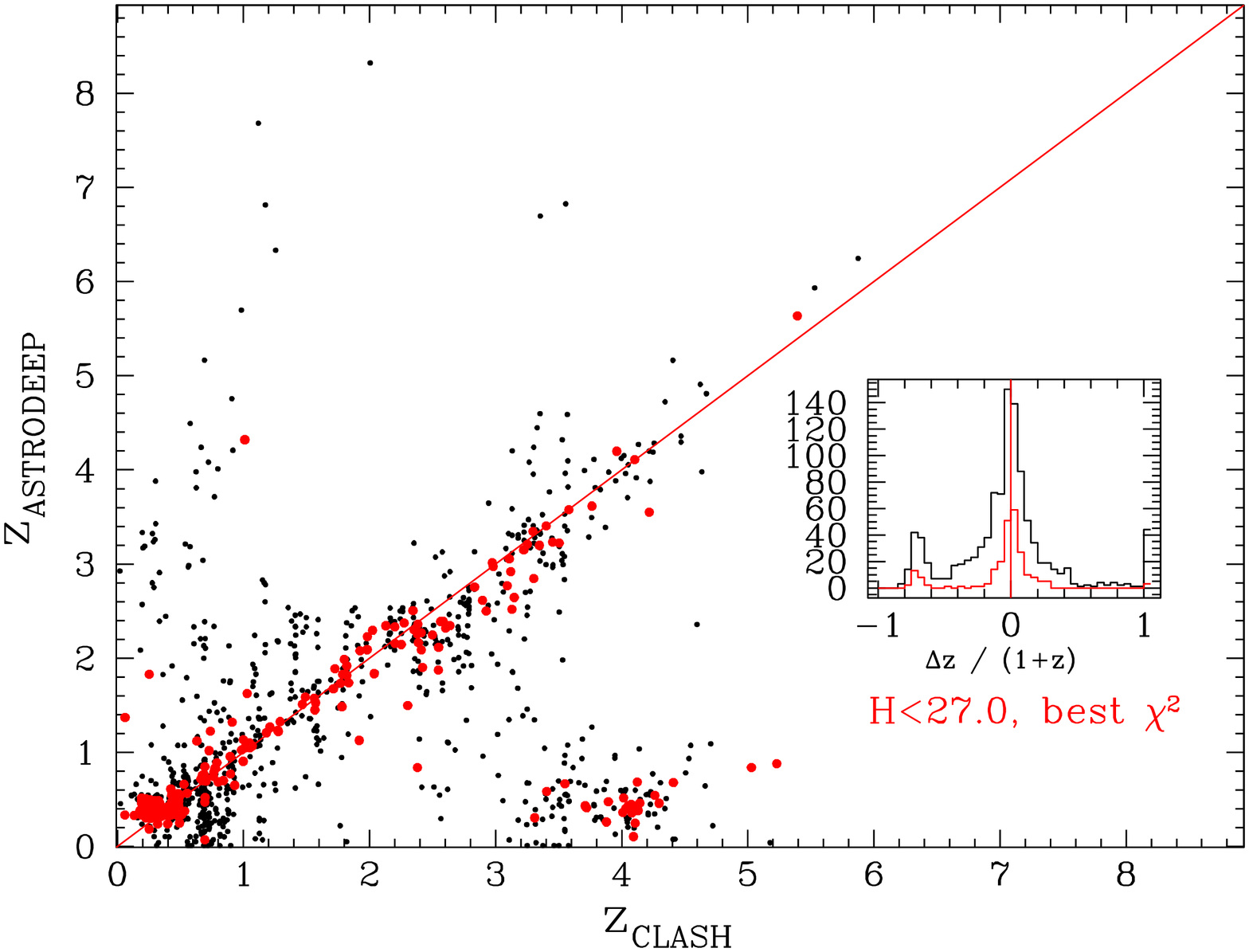}
   \includegraphics[width=6.5cm, angle=-90]{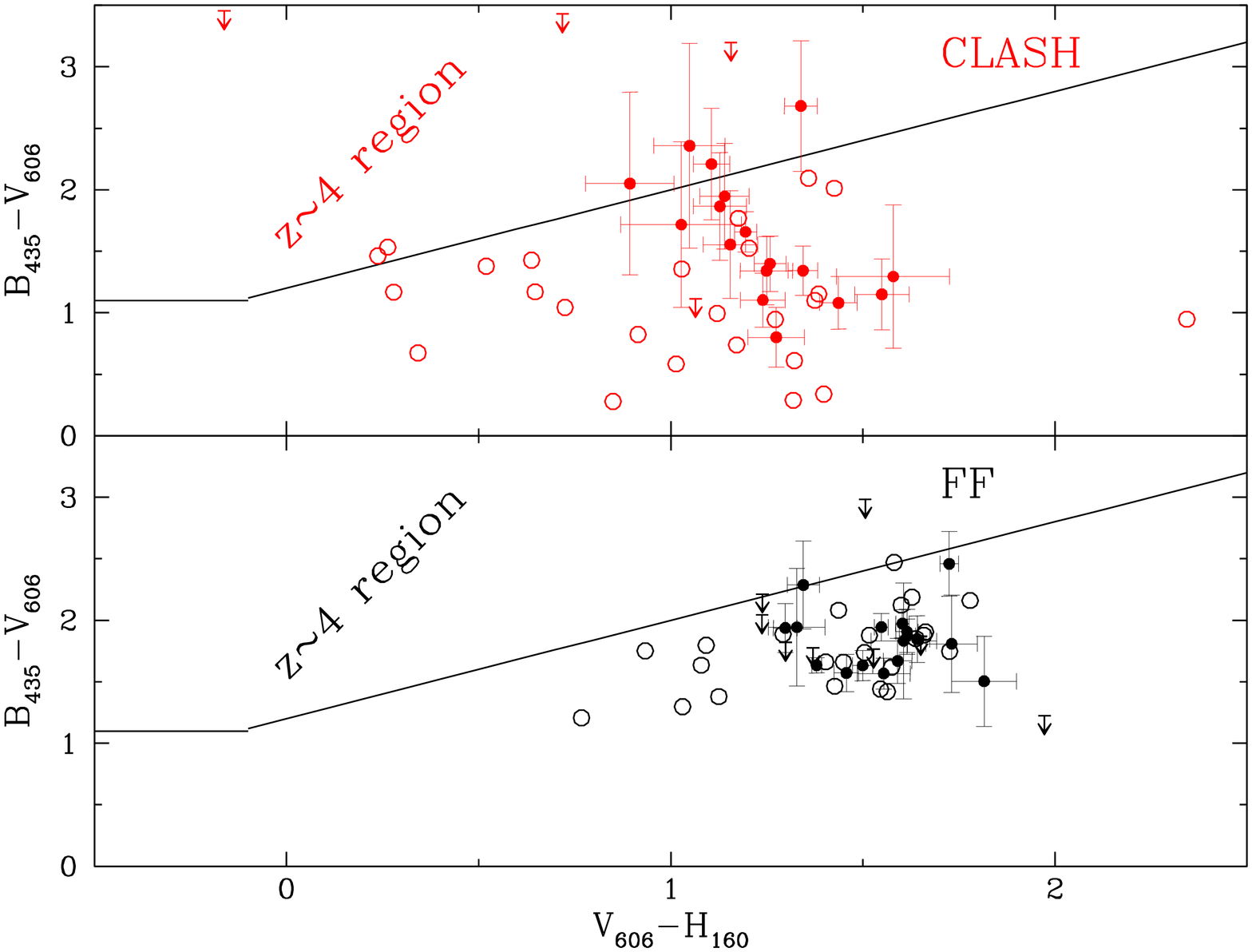}
  \caption{Top: comparison between photometric redshifts from our catalogues and those from CLASH. Red points indicate bright objects with single-peaked reliable photo-z solution according to the CLASH team (see text for details). Bottom: position in the (B-V)-(V-H) z$\sim$4 selection diagram of objects with $3.5<z_{CLASH}<4.5$ and z$_{ASTRODEEP}<$1. CLASH photometry is shown in red, photometry from the Astrodeep catalogue in black. Filled circles with errorbars represent the bright sample with reliable, single-peaked CLASH photoz.}\label{fig_clash}%
    \end{figure}

Galaxy physical properties are then computed by fitting \citet{Bruzual2003} (BC03) templates with the \verb|zphot.exe| code at the previously determined median photometric redshift. In the BC03 fit we assume exponentially declining star-formation histories with e-folding time 0.1$\leq \tau \leq$ 15.0, a \citet{Salpeter1955} inital mass function and we allow both \citet{Calzetti2000} and Small Magellanic Cloud \citep{Prevot1984} extinction laws. Absorption by the intergalactic medium (IGM) is modeled following \citet{Fan2006}. We consider the following range of physical parameters: $0.0\leq E(B-V) \leq1.1$, Age$\geq 10$Myr (defined as the onset of the star-formation episode), metallicity $Z/Z_{\odot}=0.02,0.2,1.0,2.5$. We fit all the sources both with stellar emission templates only and including the contribution from nebular continuum and line emission following \citet{Schaerer2009} under the assumption of an escape fraction of ionizing photons $f_{esc}=0.0$ \citep[see also][for details]{Castellano2014}. 

\begin{figure}[!ht]
   \centering
   \includegraphics[width=7cm, angle=-90]{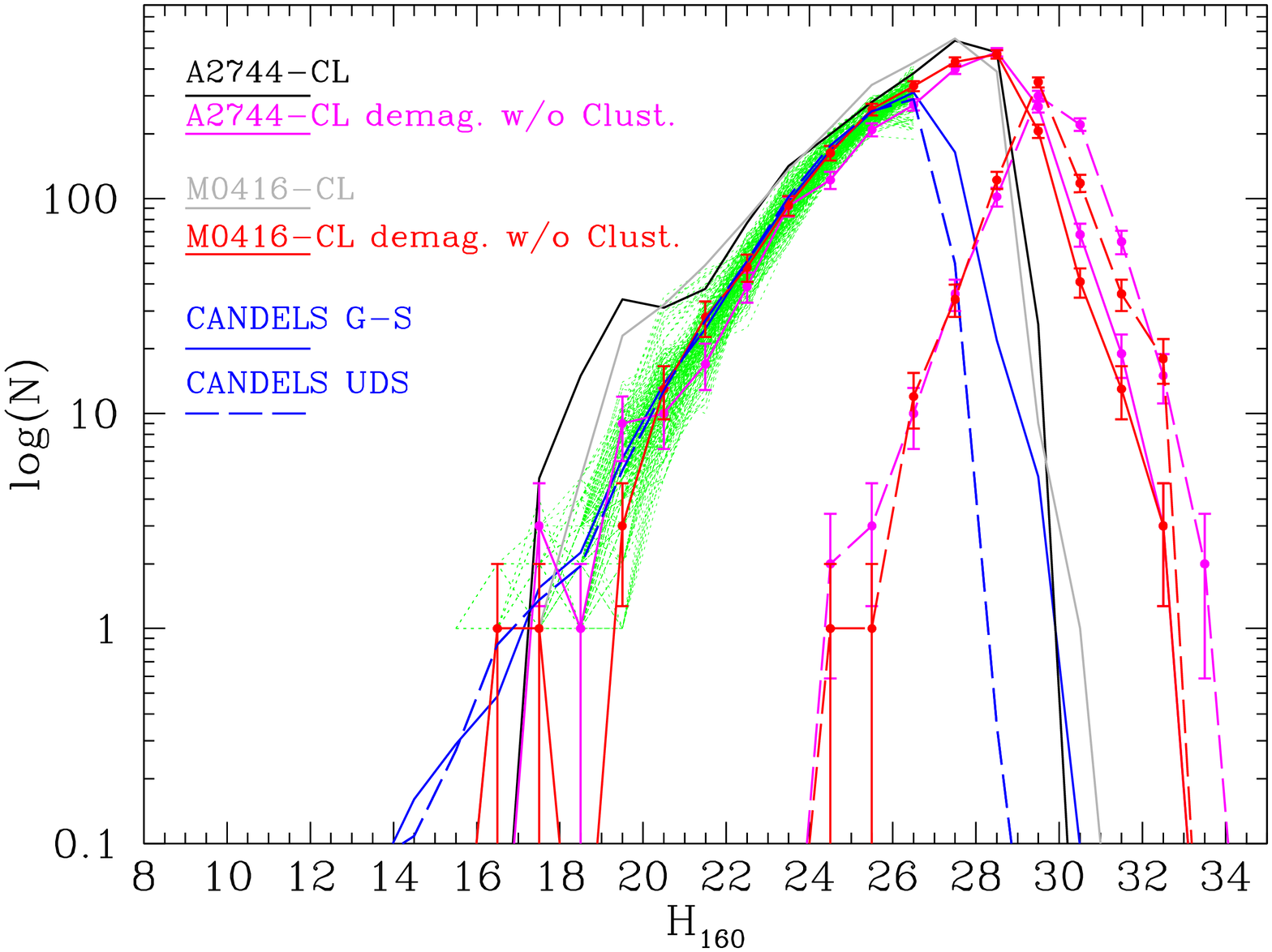}
   \caption{Demagnified (median magnification, see text for details) H160 number counts in the cluster fields. Magenta and red continuos curves refer to Abell-2744 and MACS-0416 H-detected sources respectively after excluding all objects with photoz consistent with the redshift of the clusters. Magenta and red dashed lines show the demagnified number counts of additional IR-detected sources (with S/N(H160)$>$1) in each field. As a comparison, number counts normalized to the FF area from the public CANDELS GOODS-South \citep{Guo2013} and UDS \citep{Galametz2013} catalogues are shown as continuous and dashed blue lines respectively. The green lines are number counts from randomly chosen portions of the CANDELS GOODS-South and UDS field having the same area as the FF pointings.} \label{fig_CLdemagcounts}%
\includegraphics[width=7cm, angle=-90]{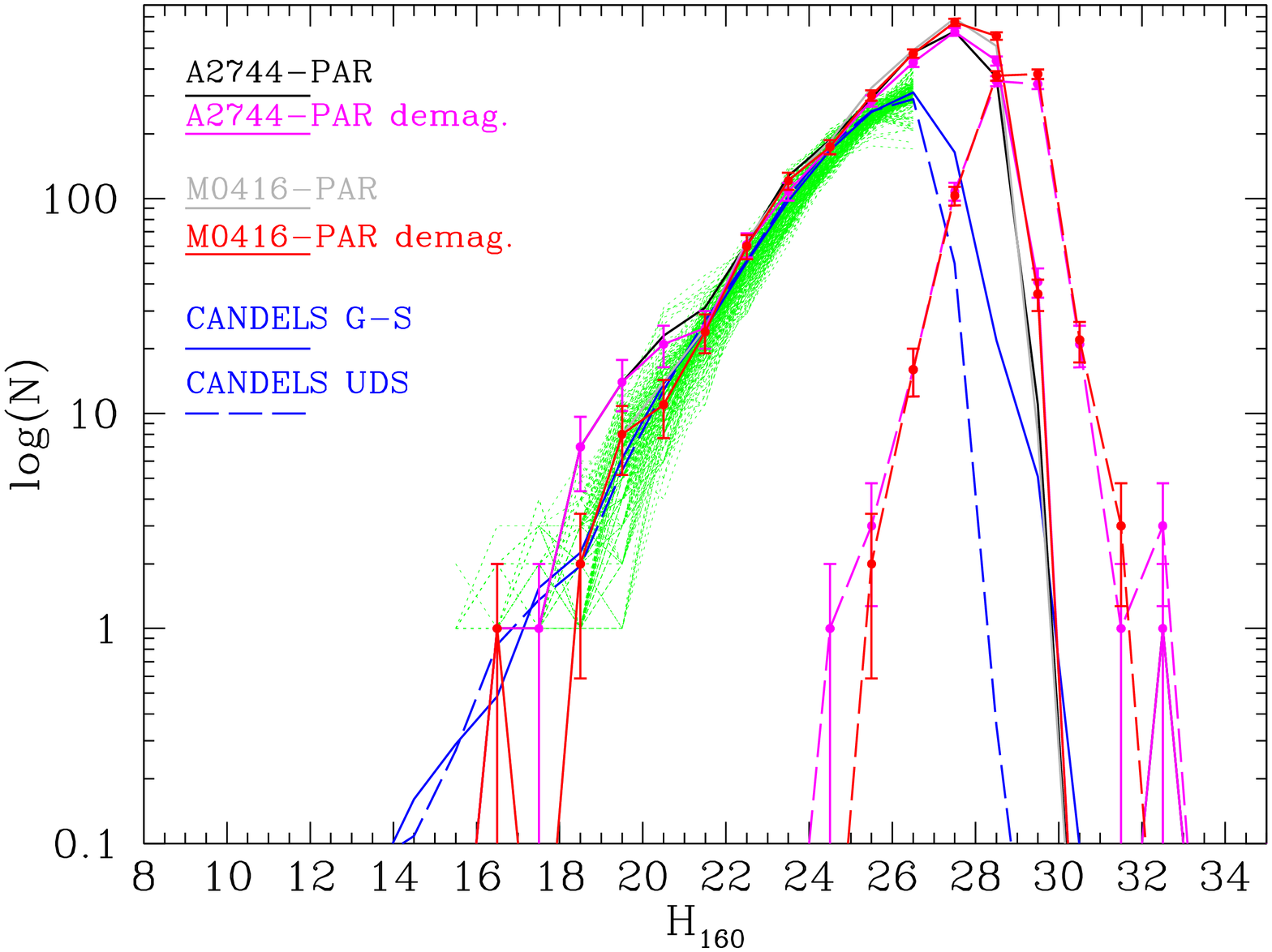}
\caption{Same as Fig.~\ref{fig_CLdemagcounts} for the Abell-2744 and MACS-0416  parallel fields.}\label{fig_PARdemagcounts}%
 \end{figure}

Photometric redshifts and rest-frame properties are determined using all 10 available bands with the following exceptions: 1) HST bands having \verb|SExtractor| \verb|FLAG|$\ge$16 and/or unphysical fluxes or uncertainties (typically truncated or problematic sources); 2) K-band and IRAC fluxes associated to a max-covariance ratio \verb|MaxCvRatio|$\ge$ 1.0 in the relevant \verb|T-PHOT| extraction indicating that the measurement is poorly reliable due to severe blending with other sources \citep[see][]{Merlin2015}. As a result, all the 10 bands are used in the fit for $\sim$65\% of the sources in the cluster pointings and for $>$90\% in the blank fields. Most of the remaining objects are fit with HST+$Ks$ photometry while one or both (for $\sim$25\% of the sources) the IRAC bands are excluded due to the large covariance. The resulting photometric redshift distributions in the four fields are shown in Fig.~\ref{fig_zdistr}, the comparisons between photometric and spectroscopic redshifts are shown in Fig.~\ref{fig_zspezbest} for the two cluster pointings. The latter is computed only on the sources with reliable photometry, i.e. excluding the areas subject to \verb|Galfit| subtraction of bright sources and having at least 5 HST bands available for computing the photometric redshifts (\verb|RELFLAG|=1, see Appendix). Following Dahlen et al. 2013 we define as outliers all objects having $|\Delta z/(1+z)|=|(z_{spec}-z_{phot})/(1+z_{spec})|\ge 0.15$. In Table~\ref{tabPHZ} we report the fraction of outliers along with average and rms of $\Delta z/(1+z)$ computed on the remaining objects. Clearly, the limited number of spectroscopic sources and their redshift distribution do not allow for an in-depth evaluation of the accuracy of photometric redshifts in all fields. It is safe to take as reference the two cluster fields having a larger spectroscopic sample that also includes high-redshift lensed galaxies, where we consistently find an accuracy $\sigma_{\Delta z/(1+z)}\sim$0.04 and $\sim$7\% fraction of outliers. We verified that the median photometric redshifts are more accurate than the individual runs when compared to spectroscopic redshifts. In the cluster fields the individual runs show similar performances with $\sigma_{\Delta z/(1+z)}\sim$0.05-0.06, $\langle\Delta z/(1+z)\rangle \gtrsim 10^{-3}$ and 8-11\% fraction of outliers.
The photometric redshift accuracy at the redshift of the clusters is comparable to the global one, implying that the redshifts presented here can be used to individuate non-spectroscopic cluster members (see also Fig.~\ref{fig_CLdemagcounts}). However, we caution against a possible tendency for a luminosity dependent behaviour of the photometric redshift offset in the M0416 cluster field, with likely cluster members at H160$>26$ having a typical offset $\Delta z = + 0.05 $, which seems at the origin of a broader and slightly shifted redshift peak in the relevant redshift distribution (bottom left panel of Fig.~\ref{fig_zdistr}). The lack of spectroscopic coverage of these sources prevents a firm conclusion in this respect.
Finally, we note that the typical photometric redshift accuracy in our Frontier Fields sample is poorer than the one achieved in the CANDELS fields \citep{Dahlen2013}, and comparable in terms of scatter and offset but with a larger fraction of outliers with respect to photozs from the 16-band CLASH photometry \citep{Jouvel2014}. In order to constrain the origin of these differences we tested the performance of one of our photoz procedures (the OAR one) on the CANDELS GOODS-South catalogue restricted to the same 10 bands available for the FF and using the same SED libraries and fitting options as for the FF. The difference in depth among the FF and the GOODS fields is not a big concern here since spectroscopic samples mostly comprise high S/N sources in both cases, such that this test effectively constrain the accuracy of our procedure on the available FF bands compared to a run on the full dataset \citep[method ``E-zphot'' in][]{Dahlen2013}. We find a 8\% fraction of outliers and a $\sigma_{\Delta z/(1+z)}$=0.045, thus an accuracy comparable to the one reached in the FFs by the OAR procedure alone ($\sigma_{\Delta z/(1+z)}\sim$0.05). On the full GOODS 19-bands catalogue we found 4.1\% outliers and a $\sigma_{\Delta z/(1+z)}$=0.037, suggesting that the lower number of bands available and narrower spectral coverage is the most relevant limiting factor in the accuracy of photometric redshifts in the Frontier Fields.

\subsection{Comparison with previous works} 

A critical aspect of the Frontier Field campaign is the investigation of lensed, intrinsically faint star-forming galaxies at very high-redshifts. While the presently released catalogue is designed to a have broader use by providing data for robustly detected sources at any redshift, it is anyway useful to compare it with the available information on high-redshift samples in the fields under analysis. To this aim we cross-correlated our catalogues (adopting a matching radius of 2PSF-FWHM=0.4 arcsec) with the samples in: \citet{Laporte2014}, \citet{Zitrin2014} \citep[see also][]{Oesch2014}, \citet{Zheng2014}, \citet{Atek2015}, \citet{Coe2015} \citet{McLeod2015}, \citet{Ishigaki2015}. The most sudied field is Abell-2744 with a total sample made up of 74 LBGs at z$\gtrsim$5. We find that 58 of these sources are present in our catalogues (of which 12 are from the additional IR-detected sample). A comparison between our photo-zs and published ones for the A2744 cluster LBGs is shown in Fig.~\ref{fig_highz} indicating a good consistency with respect to previous estimates. A similar result is found by comparing with the much smaller LBG samples from the M0416 fields \citep{Laporte2015,Coe2015,McLeod2015} and from the A2744 parallel field \citep{Ishigaki2015,McLeod2015}. We inspected the 16 LBGs  missing from our A2744 cluster catalogue and found that in 3 cases they are undetected, while 13 are very close to bright galaxies and are not deblended from them. These findings are easily explained on the basis of the detection strategy we adopted. Indeed, while the aforementioned works aimed at an ultra-deep detection of small-size and faint high-redshift sources, our catalogues have been based on a compromise between an aggressive detection which is ideal for faint objects, and the capability of avoiding over-deblending of extended lower redshift sources. Nonetheless, the recovery of most of the previously found high-z candidates with comparable photo-z estimate highlights that the general-purpose catalogues presented here are effective across a wide redshift range.

We have also compared our photometric redshift catalogue for the M0416 cluster with the one\footnote{https://archive.stsci.edu/missions/hlsp/clash/macs0416/catalogs/hst/} made available by the CLASH collaboration \citep{Postman2011}. The comparison performed on objects with robust cross-correlation between the two samples (1 match within 0.2arcsec) is shown in Fig.~\ref{fig_clash}. We separately consider bright objects having an highly-reliable photo-z in the CLASH catalogue according to the parameters released by the developers: $\chi^2\leq 1$ and a high \verb|ODDS| value (we set $>$0.8) indicating a sharply peaked unimodal redshift likelihood distribution. The agreement is remarkable with the exception of a small number of sources having z$_{phot} \lesssim$0.8 in our catalogue and  z$_{phot} \sim$ 4 in the CLASH one. We looked at their position in the z$\sim$4 color selection diagram from \citet{Castellano2012} (bottom panel in Fig.~\ref{fig_clash}) finding that their colors are indeed typical of low redshift galaxies excluded from the B-dropout selection window. While this is true when both our photometric catalogue and the CLASH one are considered, the flux uncertainty and scatter is significantly larger in the latter case further highlighting the improvementes enabled by the depth of the FF dataset and, possibly, by our catalogue building procedure that includes accurate subtraction of the foreground ICL emission.
\begin{figure}[!ht]
   \centering
   \includegraphics[width=7.0cm, angle=-90]{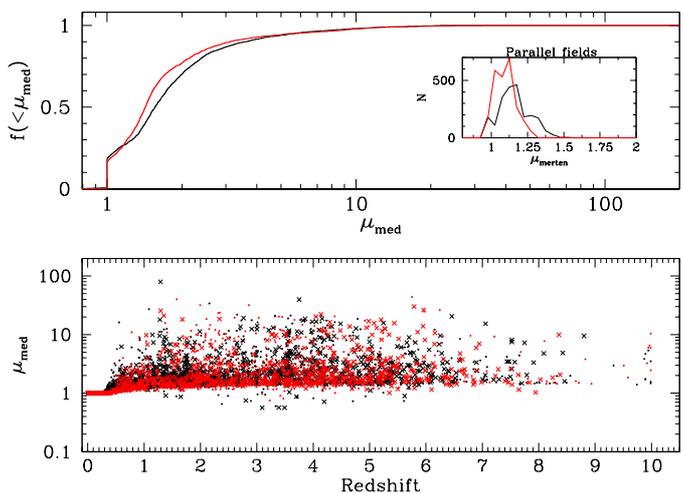}
 \caption{Top panel: cumulative distribution of the median magnification ($\mu_{med}$) values of objects in the A2744 (black) and M0416 (red) cluster fields. The inset shows the distribution of magnification values of objects in the two parallel fields according to the Merten et al. lens model.; Bottom panel: $\mu_{med}$ as a function of redshift for sources in the two fields. H-detected and IR-detected objects are drawn as filled circles and crosses respectively.}\label{fig_magnif}%
\end{figure}

\section{De-magnified number counts}\label{sect_NCOUNTS}

We use available lensing models of the two FF fields to assign magnification values to sources in our catalogues. Five of the models under consideration assume that cluster galaxies trace the cluster mass substructure: the CATS \citep[P.I. Ebeling, e.g.][]{Jauzac2014} and Sharon \citep[e.g.][]{Johnson2014} models, based on \verb|Lenstool|, the GLAFIC model \citep{Oguri2010,Ishigaki2015} and the two different parametrization (LTM and NFW) provided by the Zitrin team \citep[e.g.][]{Zitrin2013}. The three remaining models provided by P.Is Williams \citep[e.g.][]{Grillo2015}, Bradac \citep[e.g.][]{Bradac2009} and Merten \citep[e.g.][]{Merten2011}, do not assume that cluster mass is traced by its member galaxies and are instead solely constrained by lensing observables. Each team has provided shear and mass surface density maps, a detailed description of different models can be found at the FF website\footnote{http://www.stsci.edu/hst/campaigns/frontier-fields/Lensing-Models} and references therein. Among the available maps only the Merten ones cover also the parallel pointings of the fields under analysis. As a first step, we rebin the available shear and mass surface density maps to match the HST dataset pixel grid to accurately assign to each galaxy a shear ($\gamma$) and mass surface density ($\kappa$) value computed as the average in a window of 5$\times$5 pixels around its centroid. We then compute magnification as:

\begin{equation}
\mu=\frac{1}{(1- \kappa\cdot D_{zl-zp} )^2-(\gamma \cdot D_{zl-zp} )^2}
\end{equation}\label{eqmagnif}

where $D_{zl-zp}=DA(z_L,z_{phot})/DA(0,z_{phot})$, being $DA(0,z)$ the angular diameter distance to redshift $z$, $z_{phot}$ the photometric redshifts of the source and $z_L$ the redshift of the lensing cluster.
Finally, in the case of the cluster pointings where 8 different models are available, we compute a median magnification $\mu_{med}$ to take into account model-to-model variations of the lensing maps while excluding possible outlier values. The $\mu_{med}$ values as a function of redshifts and its cumulative distributions for sources in the two cluster fields are shown in Fig.~\ref{fig_magnif}. As expected, the magnification in the blank fields computed from the Merten model is nearly constant and typically low but not negligible, with median values of 15\% and 9\% in A2744-Parallel and M0416-Parallel respectively (see inset of Fig.~\ref{fig_magnif}, top panel). 
The demagnified number counts are shown in Fig.~\ref{fig_CLdemagcounts} and Fig.~\ref{fig_PARdemagcounts} for the cluster and parallel pointings respectively compared both with total number counts normalized to the FF area from the CANDELS GOODS-South \citep{Guo2013} and UDS \citep{Galametz2013} surveys, and with number counts from randomly chosen portions of the CANDELS fields having the same area as the FF pointings. At bright magnitudes the FF number counts are consistent with the CANDELS ones once magnification is taken into account and, in the case of the cluster pointings, sources with redshift compatible with being members of the A2744 and M0416 clusters (z$_{phot}$ within $\Delta$z=0.1 from the cluster redshift) are removed. At faint magnitudes the Frontier Fields cluster pointings allow us to detect sources up to $\sim$3-4 magnitudes intrinsically fainter than objects in the deepest areas of the CANDELS fields.

\begin{figure*}[!ht]
   \centering
   \includegraphics[width=12cm, angle=-90]{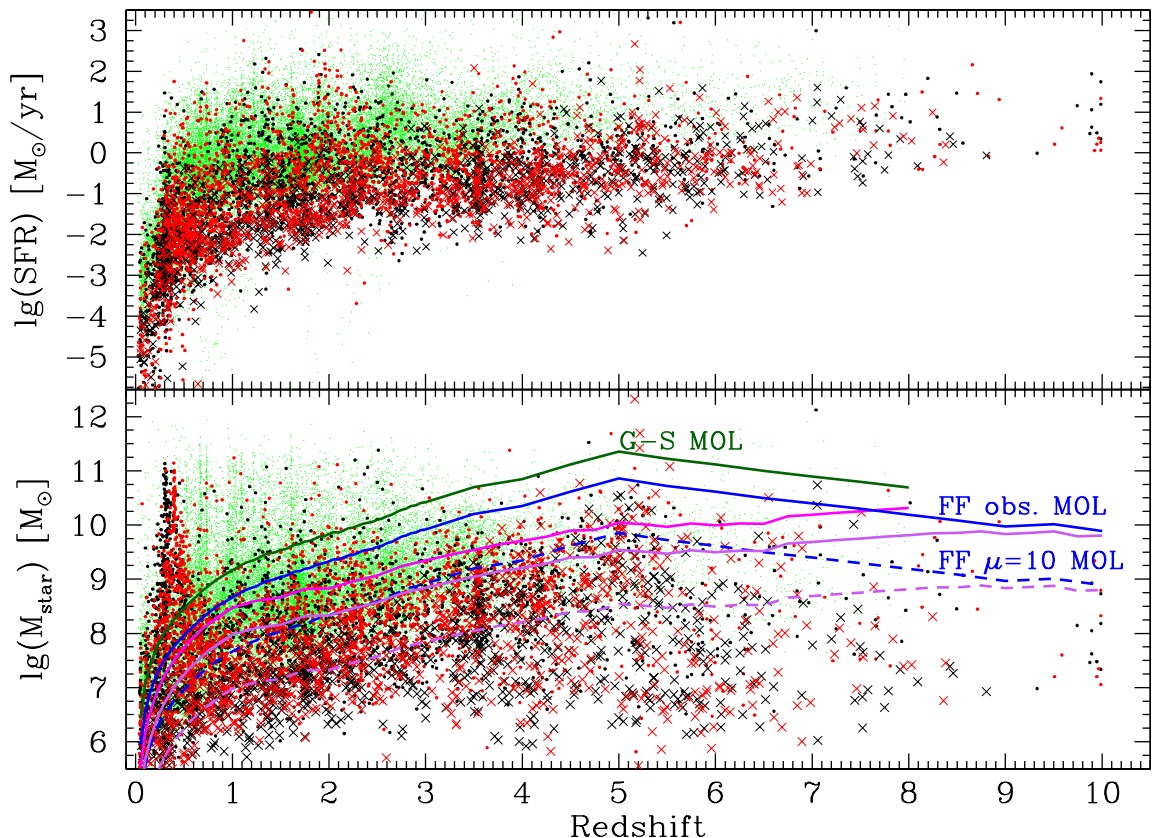}
   \caption{Top panel: Demagnified SFR as a function of redshift in the A2744 (black) and M0416 (red) clusters. Filled circles and crosses refer to H-detected and IR-detected sources respectively. For reference SFRs of objects from the CANDELS GOODS-South field are shown as green dots. Bottom panel: demagnified stellar masses, same symbols as above. The continuous and dashed blue lines show the limiting $M_{star}$ for a ``maximally old model'' at H160=27.25 (observed 90\% completeness limit) and H160=29.75 (90\% completeness limit for $\mu$=10). The continuous and dashed lines show the corresponding ``completeness corrected'' limiting $M_{star}$ (see text for details). The dark green and magenta continuous lines show the ``maximally old'' and completeness corrected limiting $M_{star}$ of CANDELS GOODS-South from \citet{Grazian2015}.}
              \label{fig_physprop}%
    \end{figure*}
\section{Rest-frame physical properties}\label{sect_RESTPROP}
The ultra deep IR observations of the FF in combination with the strong gravitational lensing effect allows one to probe stellar masses and star-formation rates at unprecedented limits. We report in Fig.~\ref{fig_physprop} the de-magnified M$_{star}$ and SFRs as a function of redshift for galaxies in the two cluster fields compared to the sample from CANDELS GOODS-South, among the most studied ``wide'' fields for investigating these properties at high-z. Intrinsic M$_{star}$ and SFR are obtained by correcting the estimates derived through SED-fitting on observed magnitudes (Sect.~\ref{sect_PHOTOZ}) considering for each source its $\mu_{med}$ value. We derive the mass completeness limits through the procedure presented in \citet{Fontana2004} which is based on the measurement of the actual distribution of the $M_{star}/L$ as a function of redshift to derive limiting stellar mass beyond the flux limit. We base our computation on the $M_{star}/L$ distributions derived by \citet{Grazian2015} from GOODS-South that provides a large and deep enough (after the inclusion of the HUDF) sample to this purpose. The strict completeness limit corresponding to H160=27.25 (90\% detection completeness for R$_{h}$=0.2 arcsec disks in the FF) is shown as a continuous blue line, the dashed one corresponding to the case of galaxies magnified by a factor $\mu$=10. These ``maximally old limits'' (MOL) are derived by considering the model with the lowest $M_{star}/L$ in our synthetic library, i.e. maximally old galaxies with formation redshift z=20, a declining SFH ($\tau$= 0.1 Gyr), E(B-V) = 0.1 and Z=0.2$Z_{\odot}$.
Clearly, the observed sample reaches lower $M_{star}$ values for less extreme galaxy populations: we show as continuous (observed limit) and dashed (for the case of $\mu$=10) purple lines the mass limits at which a completeness correction factor lower than 1.5 needs to be applied by taking into account the appropriate $M_{star}/L$ distribution \citep[see][for a detailed description of the procedure]{Fontana2004}. A comparison with the corresponding MOL and ``completeness corrected'' limits for the GOODS-South Wide field (H160$_{lim}$=26.0) taken from \citet{Grazian2015} show that the Frontier Fields clusters allow us to probe the galaxy stellar mass distribution at 0.5-1.5 dex lower masses, depending on magnification, with respect to GOODS. The inclusion of the addtional sample of IR-detected objects yield sources at $M_{star}$ as low as $10^7$-$10^8$ at high-z, although a formal derivation of completeness limits is not straightforward in this case. As shown in the top panel of Fig.~\ref{fig_physprop}, the Frontier Fields also allow to probe high-z galaxies at intrinsic SFRs $>$1dex lower than in the wide GOODS-South area, reaching 0.1-1 $M_{star}/yr$ at z$\sim$6-10.

\section{Summary and Conclusions}\label{sect_SUMMARY}
We have presented a public release of photometric redshifts and rest-frame galaxy properties from multi-wavelength photometry of the Frontier Fields A2744 and M0416 cluster and parallel pointings including both HST and deep K band and Spitzer data, as described in the companion paper Merlin et al. 2015. We have derived photometric redshifts as the median among six different estimates coming from a variety of codes and approaches (Sect.~\ref{sect_MW}). Their typical accuracy, as defined from the semi-interquartile range of the different measurements, goes from 0.05 to 0.3 at bright and faint magnitudes respectively, with less than 10\% of sources having SIQR>1 at H$<$26 and about 20\% at H$\sim$29. A comparison with available spectroscopic samples consistently show a $\sigma_{\Delta z/(1+z)}\sim$0.04 with a 7-10\% fraction of outliers. We find that the most important factor limiting the accuracy of photometric redshifts in the FF is the relatively low number of filters available compared to other surveys, such that extending the FF spectral coverage is the most promising way to improve accuracy in photometric redshifts and derived quantities. We have determined magnification values from all available lensing models on an object-by-object basis taking into account source positions and redshifts. The resulting de-magnified number counts (Sect.~\ref{sect_NCOUNTS}) are perfectly consistent with number counts from the CANDELS fields at the bright end, while reaching out to an intrinsic H$\gtrsim$32. We have shown that the Frontier Fields survey allows us to detect objects with stellar mass $M_{star} \sim 10^7$-$10^8 M_{\odot}$ and intrinsic SFRs $\sim$ 0.1-1 $M_{\odot}/yr$ at z$>$5 (Sect.~\ref{sect_RESTPROP}). Photometric redshifts, magnification values, rest-frame properties and supporting information are all made publicly available as described in the Appendix.

\begin{acknowledgements}
The research leading to these results has received funding from the European Union
Seventh Framework Programme (FP7/2007-2013) under grant agreement n. 312725. RJM, EMQ and AM acknowledge the support of the European Research Council via the award of a Consolidator Grant (PI McLure). The M0416 sppectroscopic data were based on the ESO VLT Large Programme (prog.ID 186.A-0798, PI: P. Rosati).  The financial support from PRIN-INAF 2014: ``Glittering Kaleidoscopes in the sky, the multifaceted nature and role of galaxy clusters'' (PI M. Nonino) is acknowledged. This work utilizes gravitational lensing models produced by PIs Brada\u{c}, Ebeling, Merten \& Zitrin, Sharon, and Williams funded as part of the HST Frontier Fields program conducted by STScI. STScI is operated by the Association of Universities for Research in Astronomy, Inc. under NASA contract NAS 5-26555. The lens models were obtained from the Mikulski Archive for Space Telescopes (MAST).
\end{acknowledgements}

\bibliographystyle{aa}

\begin{appendix} 
\section{Released catalogues}\label{sect_APP_CATAL}

All the catalogues and derived quantities described in this paper are publicly released and can be downloaded from the ASTRODEEP website at \texttt{http://www.astrodeep.eu/frontier-fields/}. The catalogues and images can be browsed from a dedicated interface at \texttt{http://astrodeep.u-strasbg.fr/ff/index.html}
 \\
\\
Photometric redshift catalogues contain the following information: 

\begin{itemize}
 \item  \verb|ID|: identification number in the input photometric catalogues from M16. The IR-detected objects have ID=20000+their original ID in the relevant detection catalogues and segmentation maps.

\item  \verb|ZBEST|: corresponds to the reference (median) photo-z value except when a match with a publicly available high-quality spectroscopic source is found within 1 arcsec. Sources for which the photo-z run did not converge to a solution are set to \verb|ZBEST|=-1.0.

\item  \verb|ZBEST_SIQR|: median photometric redshift uncertainty range (equal to 0 for spectroscopic sources).

\item  \verb|MAGNIF|:  median magnification (cluster fields), or magnification from the Merten model (parallel fields).

\item  \verb|ZSPECFLAG|: the value is set =1 for sources with  spectroscopic redshift, =0 otherwise.

\item  \verb|ZSPECID|: identification number of spectroscopic counterpart from public catalogues.

For Abell2744 the following convention is used: sources from Owers et al. 2011 have \verb|ZSPECID| equal to the row index in the original file; sources from Johnson et al. 2014 have \verb|ZSPECID| equal to 3000 +  row index from Table 2 in the paper, objects from the GLASS survey have \verb|ZSPECID|=10000 + original ID. 

For MACS0416 the following convention is used: sources from Ebeling, Ma  Barrett, 2014 have \verb|ZSPECID| equal to the original ID; the strongly lensed galaxies made available by STSci for FF lensing modeling \citep[from ][]{Grillo2015,Christensen2012} have \verb|ZSPECID|=3000 + row index from the original file, objects from the GLASS survey have \verb|ZSPECID|=10000 + original ID.

The value is -1 for sources with no spectroscopic counterpart.

\item \verb|Chi2|: $\chi^2$ of the SED fitting with stellar only templates at redshift fixed to ZBEST.

\item  \verb|MSTAR|, \verb|MSTAR_MIN|, \verb|MSTAR_MAX|: stellar mass in units of $10^{9} M_{\odot}$ (assuming Salpeter IMF) and relevant uncertainty range. Uncertainties on physical parameters are defined from the range where P($\chi ^2$)$>$32\% estimated in a $\Delta$z=0.2 redshift bin around the reference photometric redshift.

\item  \verb|SFR|, \verb|SFR_MIN|, \verb|SFR_MAX|: star-formation rate ($M_{\odot}/yr$) and relevant uncertainty range.

\item \verb|Chi2_NEB|: $\chi^2$ of the SED fitting with stellar plus nebular models at redshift fixed to ZBEST.

\item  \verb|MSTAR_NEB|, \verb|MSTAR_MIN_NEB|, \verb|MSTAR_MAX_NEB|: stellar mass ($10^{9} M_{\odot}$) estimated from stellar plus nebular fits.

\item  \verb|SFR_NEB|, \verb|SFR_MIN_NEB|, \verb|SFR_MAX_NEB|: star-formation rate ($M_{\odot}/yr$) estimated from the stellar plus nebular fits.

\item  \verb|RELFLAG|: This flag is meant to provide a combined indication of the robustness of photometric and photo-z estimates. Sources with \verb|RELFLAG|=1 have enough reliable photometric information for estimating photometric-redshifts. Instead, the value is =0 for sources either: falling close to the border of the images; close to strong residual features of the Galfit image pre-processing; found to be spurious (mostly stellar spikes) from visual inspection; having SExtractor \verb|FLAG|>=16; having unphysical flux in the detection band; having less than 5 HST bands with reliable flux measurement available for photo-z procedures.

\end{itemize}

\end{appendix}

\end{document}